\documentclass[
aps,nofootinbib,showpacs,showkeys,preprint
tightenlines,preprintnumbers,] {revtex4}

\usepackage{epsf,epsfig,subfigure,graphicx,amsmath,amssymb,mathtools}
\usepackage{color}
\newcommand{\dis}[1]{\begin{equation}\begin{split}#1\end{split}\end{equation}}
\newcommand{\ie}{{\it i.e.~}}
\newcommand{\etal}{{\it et al.\,}}

\newcommand{\gev}{\,\textrm{GeV}}

\newcommand{\Mp}{M_{\rm P}}
\newcommand{\Mpt}{$M_{\rm P}$}
\newcommand{\Mg}{{M_{\rm GUT}}}
\newcommand{\Mgt}{$M_{\rm GUT}$}

\begin{document}
\draft

\title{\Large\bf   Numerical calcultaion of $e$-fold number from hilltop with two inflatons}

\author{ Jihn E.  Kim$^{(a)}$ and Doh Young Mo$^{(b)}$}
\affiliation
{
$^{(a)}$Department of Physics, Kyung Hee University, 26 Gyungheedaero, Dongdaemun-Gu, Seoul 130-701, Republic of Korea\\
$^{(b)}$Department of Physics and Astronomy, Seoul National University, 1 Gwanakro, Gwanak-Gu, Seoul, 151-747, Republic of Korea
}

\begin{abstract} 
We consider a cosmological inflation with two inflatons, $\phi$ and $X$. The inflation potential is a hilltop form in the $\phi$ space and $X$ is a sideway down-fall field in the region $\phi\ge m$. In this model, we calculate the $e$-fold number numerically starting from the intitial point which is in the vicinity of  the hilltop point. Firstly, by varying parameters and initial conditions the inflaton path field $\varphi$ for the BICEP2 point,  $\varphi_{\rm BCP2}$, is found to give $ n_s\simeq 0.96$ and $r\simeq 0.16$. Next, from the point  $\varphi_{\rm BCP2}$ to the end of inflation the $e$-fold number is obtained. We find a reasonable set of parameters allowing the
 $e$-fold number in the range $50-60$.

\end{abstract}

\maketitle


\section{Introduction}\label{sec:Introduction}

The recent BICEP2 announcement on the tensor-to-scalar ratio at $r = 0.2^{+0.07}_{-0.05}$ ($r=0.16^{+0.06}_{-0.05}$ after the elimination of the dust background \cite{BICEP2I})  has openned up a new era in the inflationary cosmology.   Although there still exists a possibility that most of it is due to the dust background if one extrapolates the Planck dust analysis \cite{PlunckDust14}, it is so an important discovery even with its confirmation at the level of a tenth\footnote{It might be possible in the near future by BICEP3, Keck Array, and liteBIRD experiments.}  of the announcement that a particular attention must paid to this announcement.  Since the temperature perturbation $\Delta T/T$ obtained from the scalar perturbation is at the level of  $10^{-5}$, the tensor perturbation of order\footnote{The factor $16\pi^2$ is just a proportional factor for an order of an estimate to give the BICEP2 value $r\approx 0.16$ for $\Mg\approx 2\times 10^{16\,}\gev$.}  $16\pi^2 \Mg^4/\Mp^4\approx 10^{-10}-10^{-6}$ (for $\Mg\approx (2-20)\times 10^{15\,}\gev$) will lead to $r\approx O(10^{-5}-10^{-1})$. Namely, even the confirmation of $r$ at the level $10^{-5}$ implies that there was the inflationary potential energy density of order $ (2\times 10^{15\,}\gev)^4$. It is a grand unification (GUT) scale, and hence the GUT scale inflation needs to be scrutinized.

Inflationary models need some fine-tuning to fit to the observed valuess of  $\Delta T/T$, non-Gaussianity, $n_s, r$, etc. For example, the $\frac12 m^2\phi^2$ single field chaotic inflation \cite{LindeChaotic} needs the mass parameter at the level of $10^{-10}$ in units of the reduced Planck mass $\Mp^2 $ and the inflaton field excursion $\langle \phi_{\rm inflaton}\rangle$ needs to be trans-Planckian, \ie $\ge 15\Mp$ \cite{Lyth97}. Why we keep only the  $\frac12 m^2\phi^2$ term is a kind of an extreme fine-tuning problem \cite{Lyth14,KimHilltop14}. The $\frac12 m^2\phi^2$ term is not the only term, but also a host of terms is allowed with  the discrete symmetry allowed in string compactification \cite{Kimplb13}. This will in general leads to a hilltop potential.

For theories toward inflation, however, physics beyond the Standard Model  is not firmly established yet. So, it is a simple way to work with theoretically attractive ideas among which the leading ones are GUTs and supersymmetry (SUSY). After the discovery of Higgs boson, the Higgs portal to high (GUT and/or intermediate) energy scale
is another widely used method. So, we consider a simple scenario of  three scales, the Planck scale $\Mp\simeq 2.43\times 10^{18\,}\gev$,  the GUT scale $\Mg\simeq 2\times 10^{16\,}\gev$, and the electroweak scale $v_{\rm ew}\simeq 247\,\gev$.
The intermediate axion scale and the SUSY breaking scale can be considered as derived ones from these. In this way, we can introduce the axion window scale \cite{Kim84,KimNilles84,Kim13} by assuming a Peccei-Quinn symmetry at a GUT scale  \cite{Kim79} which is broken at the intermediate scale. On the other hand, it was noted that gravitational interactions invalidate the Peccei-Quinn symmetry \cite{PQ77} broken at the intermediate scale, which
seemed to have excluded the invisible axion idea \cite{GravSpoil92}. But, the global symmetry breaking terms including the baryon number violating operators can be sufficiently suppressed in some string compactification models  \cite{stringglobal06}. When this sufficient suppression of the global symmetry breaking terms is realized in string theory, we are free from the gravity spoil of global symmetries because string theory already contains a consistent gravity theory with certain exact discrete symmetries \cite{Kimplb13}. If the global symmetry breaking terms appears at a sufficiently high order, then even the dark energy scale of O($10^{-47}\gev^4$) can be interpreted in some string compactifications with a high order of discreteness \cite{KimNilles14,KimJKPS14}.

Also, the inflaton field is not free from the gravity spoil of global symmetries.  For the case of  inflaton potentials,  therefore, one only considers the terms implied by string theory, with two leading scales of \Mpt~ and \Mgt.
Since a large trans-Planckian excursion is needed for the case of single inflaton $ \phi_{\rm inflaton}$, one worries about the huge energy density  $ \phi_{\rm inflaton}^n$ at the trans-Planckian field value. This has led to the natural inflation idea \cite{Freese90} such that higher order terms are arranged to cut off huge energy down to $\Lambda_{\rm GUT}^4$ by nonperturbative effects of a confining gauge group. It is a GUT-scale axion inflation. However, the axion decay constant is of order \Mgt. A large axion decay constant may be introduced by a small quartic coupling constant $\tilde\lambda$,
\dis{
V=\frac1{4!}\tilde\lambda(|\phi|^2-f^2)^2.\nonumber
 }
The mass parameter in this theory is $m^2=\tilde\lambda f^2/6$ which is interpreted as a GUT scale. Then, $f$ can be trans-Planckian of order $>10\Mp$ for $\tilde\lambda< 10^{-5}$. However with this small $\tilde\lambda $,  `inflation' is already in the radial direction, and the angle direction of the natural inflation is not needed.
To realize the trans-Planckian axion decay constant in the natural inflation, therefore, we require (i) the radial field decays rapidly with $\tilde\lambda \gg 10^{-5}$, and (ii) more confining forces at the GUT scale are introduced as done by Kim, Nilles and Pelsoso (KNP) \cite{KNP05}. This KNP model already introduces two axions and therefore the hilltop inflation with two inflatons is not much worse than the KNP model. In fact, the hilltop inflation with two inflatons was suggested in Ref. \cite{KimHilltop14} and this paper is a numerical study on the feature envisioned there. Here, we extend the single scalar field hilltop inflation to two scalar fields inflation and perform a numerical study of inflation starting from an initial inflation point near the hilltop. The purpose of introducing the second inflaton is to have a significant second derivative of the inflaton potential so that a large $r$ can result to fit to a value of O(0.1) by a positive $V^{\prime\prime}$, since the $n_s$ and $r$  relation to the slow-roll parameter $\eta\,(=\Mp^2V^{\prime\prime}/V)$ is $n_s=1-\frac38 r+2\eta$. To fit to $n_s\simeq 0.96$ with a large $r$, one needs a large positive second derivative of $V$.

In Sec. \ref{sec:Model}, we present a two inflatons model. In Sec. \ref{sec:NumStudy}, we numerically solve the evolution equations of these two inflatons and present a scatter plot of the parameters within the required interval of the e-fold number and compared them to the Planck and BICEP2 data. Sec. \ref{sec:Conclusion} is a conclusion.

\section{Model}\label{sec:Model}

The one inflaton hilltop potential with the symmetry $\phi \rightarrow - \phi$, satisfying  three conditions $V'(0)=0$, $V(f_{\rm DE}) = 0$, $V'(f_{\rm DE}) = 0$, can be parametrized as
\dis{
V = \frac{\lambda M_{\rm P}^4}{4!} (\phi^2 - f_{\rm DE}^2)^2 \equiv \frac{\lambda}{4!} (\phi^2 - f_{\rm DE}^2 )^2.\nonumber
}
The dimensionful variables will be expressed in units of \Mpt~in the following discussion, and we will set $\Mp=1$ if not stated explicitly.
The  first and second derivative are
\dis{
V_{,\phi}      = \frac{\lambda \phi}{6} ( \phi^2 - f_{\rm DE}^2 ), ~~
V_{,\phi \phi} = \frac{\lambda}{2} (\phi^2 - \frac{1}{3} f_{\rm DE}^2).\nonumber
}

Then, $r$ is given by
\dis{
r = 8 \left( \frac{\frac{\lambda \phi}{6} ( \phi^2 - f_{\rm DE}^2 ) }{\frac{\lambda}{4!} (\phi^2 - f_{\rm DE}^2)^2} \right)^2 = \frac{128 \phi^2}{(\phi^2 - f_{\rm DE}^2)^2}.\nonumber
}
Since the reported central value of BICEP2 is $r=0.2$ without the background dust elimination, the VEV of $\phi$ at the BICEP2 observation point is determined as $\phi_{\rm BCP2} = \pm \sqrt{ f_{\rm DE}^2 - 8 \sqrt{10} \sqrt{f_{\rm DE}^2 + 160} + 320}$. And, $\eta$ is estimated as
\dis{
\eta (\phi)=
 \frac{\frac{\lambda}{2} (\phi^2 - \frac{1}{3} f_{\rm DE}^2) }{ \frac{\lambda}{4!} (\phi^2 - f_{\rm DE}^2)^2 }  = \frac{12 (\phi^2 - \frac{1}{3} f_{\rm DE}^2)}{(\phi^2 - f_{\rm DE}^2)^2 }.\nonumber
}
Substituting $\phi_{\rm BCP2}$,
\dis{
\eta(\phi_{\rm BCP2}) = \frac{0.125 f_{\rm DE}^2 - \frac{ 3 \sqrt{10 f_{\rm DE}^2 + 1600}}{2} + 60}{(\sqrt{10} \sqrt{f_{\rm DE}^2 + 160} - 40)^2} .\nonumber
}
The inflaton $\phi$ eventually converges to the ground value $f_{\rm DE}$. Since the $\phi$ value in the expression of $\eta$ is taken at the BICEP2 observation point, $\phi$ tends to be much smaller than  $f_{\rm DE}$ to give the $e$-fold number 55. In this case, $\eta$ is negative and $n_s=0.96$ cannot be obtained. Even if $\phi_{\rm BCP2}$ is comparable to  $f_{\rm DE}$ to allow a positive $\eta$, then  $f_{\rm DE}$ must be at least around 40.  Since a rough estimation of $\eta$ is O($1/ f_{\rm DE}^2$), $\eta$ is O(0.001) for  $f_{\rm DE}=40$. Thus, to raise $n_s$ from $1-\frac38 r=0.92$ to $n_s=0.96$, we need $r\approx 0.02$ because $n_s = 1 - \frac{3}{8}r + 2 \eta$. Thus, to fit to the large value of BICEP2,  $r\approx 0.2$, there is a need to introduce another inflaton.  

Therefore, let us extend the single field hilltop inflation case to the one with two inflatons $\phi$ and $X$, with the following potential,
\dis{
V = \frac{\lambda_{\phi}}{4!} ( \phi^2 - f_{\rm DE}^2 )^2 + \frac{ \lambda_{X}}{4!} \left( X^2 - \gamma [\phi^2 - m^2 ] \right)^2, \label{eq:HilltopTwo}
}
where $\lambda_{\phi}, \lambda_{X}, f_{\rm DE},$ and $ m $ are taken to be positive. Since the inflation path chooses $X=0$
at $\phi=0$ up to $\phi=\pm m$, $X$ is mimicking the waterfall field in the hybrid inflation with the superpotential given in \cite{Kim84}. The second derivative with respect to $X$ is designed to give a positive $\eta$, which is the reason that it is called `chaoton'.
Since  the large field value $\varphi$ (= inflation path field) experiences the large Hubble friction, it prevents an effective ending of inflation. 
Here, we point out what usually happens in the large field inflation scenario. The equation of motion for the inflaton is given by
 \dis{
 \ddot{\phi} + 3 H \dot{\phi} + V_{, \phi} = 0,\label{eq:evolution}
 }
where the Hubble parameter $H$ is
\dis{
  H^2 = \frac{1}{3} \left( V + \frac{1}{2} \dot{\phi}^2  \right) .\label{eq:Hubble}
 }
The Hubble friction term is large if the field value is large. In this case, the field velocity reaches its terminal velocity,
 $\dot{\phi}_{\rm ter} = -V_{, \phi} / 3H$. For the chaotic inflation with $V = \frac{1}{2} m^2 \phi^2$, $\dot{\phi}_{\rm ter}\simeq -m \sqrt{\frac{2}{3}}$. Equating kinetic energy of terminal field velocity to the potential energy, $\frac{1}{2}\dot{\phi}_{\rm ter}^2 = \frac{1}{2} m^2 \phi_{\rm end}^2$, we obtain $\phi_{\rm end} = \frac{2}{3}$. We can conclude that if the inflation starts from sufficiently large initial field value, the end point for the inflation is fixed. This behavior is quite common in any large field inflationary scenario.  Thus, another mechanism is needed to end the inflationary period. So, to end the inflation effectively when $\varphi$ passes $\varphi_{\rm max}$, we introduce the following interaction with a new field $Y$,
\dis{
\mathcal{L}_{\rm inflation~end} \ni -\theta( \varphi - \varphi_{\rm max} ) Y^2 \tilde{m}^2 \frac{| \varphi |}{\Mp}.
}
$Y$ decays to SM particles and reheats the Universe by
\dis{
\mathcal{L}_{\rm {\it Y}\,decay}  \ni \frac{Y}{\Mp}\, \mathcal{L}_{\rm SM} .
}
These terms are helpful to have appropriate $e$-fold numbes.

\section{Inflation path}\label{sec:NumStudy}

\begin{figure}[!t]
\begin{center}
\includegraphics[width=0.85\linewidth]{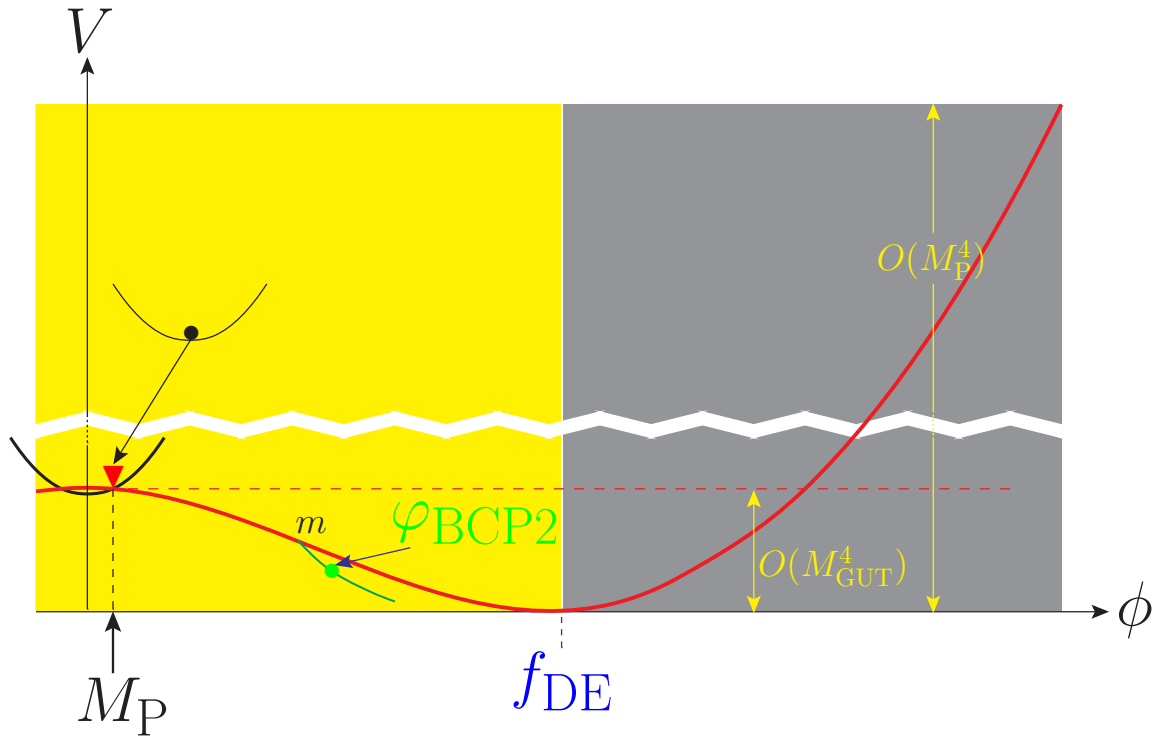}
\end{center}
\caption{The $U(1)_{\rm de}$-hilltop inflation. The red inverted-triangle is the beginning of inflation and the green bullet is at $\varphi_{\rm BCP2}$. } \label{fig:Hilltop}
\end{figure}

This feature on the inflation will be implemented in a way that  the inflation just ends at some point without meeting the inflation ending condition of the kinetic energy of inflaton becoming comparable to the potential energy. So to fit to the BICEP2 vacuum energy, there are four free parameters we can dial along the inflation path: (1) the initial point $\phi_{\rm init}$ to start the inflation where $\phi_{\rm init}$ is a bit away from the true hilltop, (2) the initial velocity $\dot X_{\rm init}=\dot X({t_{\rm init}})$, to make the inflaton roll down to the $X$ direction when $\phi$ reaches the point $\phi(t_{\rm init})$, (3) the BICEP2 point $\varphi_{\rm BCP2}$ where BICEP2 collaboration claims that they have found the large B-mode polarization on CMB, and (4) the end point of inflation $\varphi_{\rm end}\simeq \phi_{\rm max}$. It is depicted in Fig. \ref{fig:Hilltop}.

We proceed to solve Eq. (\ref{eq:HilltopTwo}). The GUT scale is $\Mg\sim 10^{-2}$, thus
the order of the inflation potential is roughly $10^{-8}$. This determines the order of couplings $\lambda_{\phi}$, $\lambda_{X}$
in following way,
\dis{
\lambda_{\phi} = \mathcal{O}(10^{-8})/f_{\rm DE}^4, ~
\lambda_{X}    = \mathcal{O}(10^{-8})/f_{\rm DE}^4.}
The parameter $ \gamma $ in Eq. (\ref{eq:HilltopTwo}) describes the relative scale for $X$ to be separated from the $\phi$ direction.
 The minimum of the potential is at
\dis{
\langle \phi \rangle = \pm f_{\rm DE},~~
\langle X \rangle    = \pm \sqrt{ \gamma (f_{\rm DE}^2 - m^2 )}.}

The dynamics of the $\phi$ inflaton is governed by  Eq. (\ref{eq:HilltopTwo}). Here, we introduce the boundary condition near the hilltop,
\dis{
\textrm{BC0:  }  \phi(0)=\phi_i,~~ \dot\phi(0)=0.\label{eq:BCatHill}
}
At the point $m$ of Fig. \ref{fig:Hilltop}, we choose the following for the numerical calculation,
\dis{
\textrm{BCm:  } \phi(t_m)=\phi_m=m,~~ X(t_m)=0,~~ \dot X(t_m)>0.\label{eq:BC}
}
As the inflaton path reaches the point $m$ of Fig. \ref{fig:Hilltop}, quantum fluctuations of $X$ can be considered and an effective separation of the inflaton path $\varphi$ from the initial $\phi$ direction can be considered. Here, this feature is mimicked by the boundary condition (\ref{eq:BC}), and take  $\dot X(t_m)$ as an arbitrary number. A similar quantum fluctuation in the beginning of inflation is mimicked by $\phi_i>0$ of Eq. (\ref{eq:BCatHill}). We can try all nonvanishing numbers for $\phi_i, \dot\phi(0), X(t_m),$ and $\dot X(t_m)$, but here we take a simple case of just varying  $\phi_i$ and $\dot X(t_m)$, which will show the key feature of the inflation path leading to a large $e$-fold number. 

\begin{figure}[!t]
\begin{center}
\includegraphics[width=0.75\linewidth]{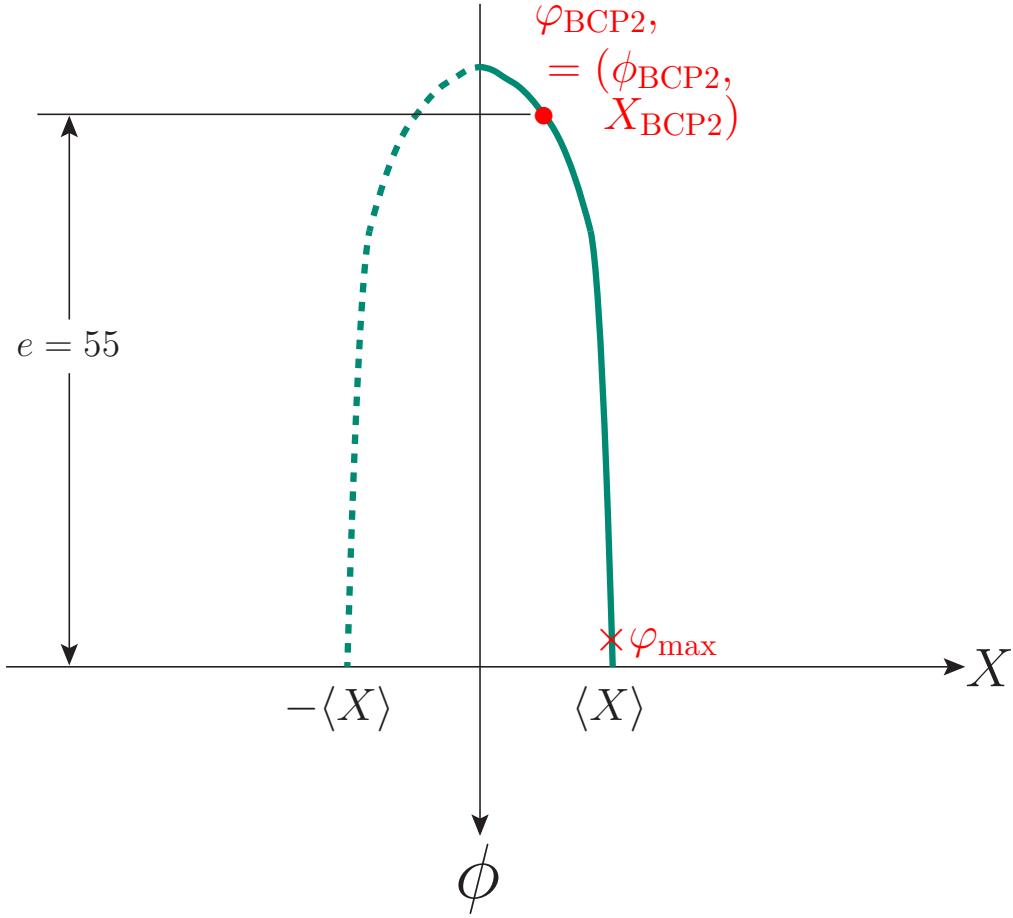}
\end{center}
\caption{The inflaton $\varphi$ path.   } \label{fig:Path}
\end{figure}

After $\phi$ passes the point $\phi( t_m ) = \phi_m$, the inflaton field moves on $\phi-X$ plane,
which is denoted as $\varphi$. A schematic behavior of the $\phi\to\varphi$ trajectory is given in Fig. \ref{fig:Path}.
 The evolution of $\phi(t)$ is calculated in two steps for the interval  $t=[0, t_{\rm BCP2}]$, and for the interval $t= [t_{\rm BCP2},\infty]$. We determine $t_{\rm BCP2}$ by varying parameters of Eq. (\ref{eq:HilltopTwo}) and the boundary conditions at the hilltop so that $r$ is in the region $[0.1,0.3]$ and $n_s$ is in the region $[0.94,0.98]$. $r$ and $n_s$ are given by
 \dis{
 r =8\left( \frac{V'_{[\varphi-{\rm dir}]}}{V}\right)^2,~~
  n_s=1-\frac38 r+2\eta,
 }
where
 \dis{
 \eta=\left( \frac{V''_{[\varphi-{\rm dir}]}}{V} \right) .
 }
Here,  $\varphi$ is the inflaton path field in the $(\phi, X)$ plane.
Partial derivatives with respect to $\phi$ and $X$ are,
\dis{
&24V_{,\phi} = 4 \lambda_{\phi} \phi (\phi^2 - f_{\rm DE}^2 ) - 4  \gamma  \lambda_{X} \phi ( X^2 -  \gamma (\phi^2 - m^2)), \\
&24V_{,\phi \phi}  = 4 \lambda_{\phi} (\phi^2 - f_{\rm DE}^2 ) + 8 \lambda_{\phi} \phi^2\\
&\quad\quad~~ - 4 \lambda_X  \gamma (X^2 -  \gamma (\phi^2 - m^2 ) ) + 8 \lambda_{X}  \gamma ^2 \phi^2, \\
&24V_{,X} = 4 \lambda_X X (X^2 -  \gamma  (\phi^2 - m^2)), \\
&24V_{,XX} = 4 \lambda_X (X^2 -  \gamma  (\phi^2 - m^2)) + 8 \lambda_X X^2, \\
&24V_{,\phi X} = -8 \lambda_X  \gamma X \phi ( X^2 -  \gamma (\phi^2 -m^2 )) = V_{,X \phi}.
}
With the information on $\varphi(t)$, the first derivative with respect to $\varphi$ direction
has the following relation,
\dis{
V'_{[\varphi-{\rm dir}]} &\equiv \nabla _{,\varphi}V= \left[ V_{,\phi} \hat{\phi} + V_{,X} \hat{X} \right] \cdot \left[ (\dot{\varphi}_{\phi} / \dot{\varphi}) \hat{\phi} + (\dot{\varphi}_{X} / \dot{\varphi}) \hat{X}   \right] \\
                         &= V_{,\phi} (\dot{\varphi}_{\phi} / \dot{\varphi}) + V_{,X} (\dot{\varphi}_{X} / \dot{\varphi})  = V_{,\phi} \frac{d \varphi_{\phi}}{d \varphi} + V_{,X} \frac{d \varphi_{X}}{d \varphi}\\
                      &   \equiv V_{,\phi} \Delta \widehat{\varphi}_{\phi} + V_{,X} \Delta \widehat{\varphi}_{X}.\label{eq:onederi}
}
Since $\dot{\varphi}(t)$ is a constant for a given $t$, the second derivative is simply given by:
\dis{
V&''_{[\varphi-{\rm dir}]}  \equiv \nabla _{,\varphi\varphi}V=  \Big[ \left(V_{,\phi \phi} \frac{\dot{\varphi}_{\phi} }{ \dot{\varphi}}
                           + V_{,X \phi}\frac{\dot{\varphi}_{X} }{ \dot{\varphi}}  \right) \hat{\phi}+ \Big(  V_{,\phi X} \frac{\dot{\varphi}_{\phi}}{ \dot{\varphi}}
                           + V_{,XX} \frac{\dot{\varphi}_{X} }{ \dot{\varphi}} \Big) \hat{X} \Big]
                           \cdot \left[ \frac{\dot{\varphi}_{\phi}}{\dot{\varphi}} \hat{\phi} +
                            \frac{\dot{\varphi}_{X}}{\dot{\varphi}}\hat{X}\right] \\
                          &= V_{,\phi \phi} \frac{\dot{\varphi}_{\phi}}{\dot{\varphi}}
                          \frac{\dot{\varphi}_{\phi} }{ \dot{\varphi}}+ V_{,X \phi} \frac{\dot{\varphi}_{X}}{
                           \dot{\varphi}} \frac{\dot{\varphi}_{\phi} }{\dot{\varphi}}+ V_{,\phi X} \frac{\dot{\varphi}_{\phi} }{ \dot{\varphi}}
                          \frac{\dot{\varphi}_{X} }{\dot{\varphi}}
                           + V_{,XX} \frac{\dot{\varphi}_{X}}{ \dot{\varphi}}
                           \frac{\dot{\varphi}_{X}}{ \dot{\varphi}} \\
                          &\equiv V_{,\phi \phi} \Delta \widehat{\varphi}_{\phi} \Delta
                           \widehat{\varphi}_{\phi}
                           + 2 V_{,\phi X} \Delta \widehat{\varphi}_{X} \Delta \widehat{\varphi}_{\phi} + V_{,XX} \Delta \widehat{\varphi}_{X} \Delta \widehat{\varphi}_{X}.\label{eq:twoderi}
}
Let us introduce the velocity vector of $\varphi$ as
\dis{
\frac{\dot{\varphi}_{\phi} }{ \dot{\varphi}} = \cos \theta ,~~
\frac{\dot{\varphi}_{X} }{ \dot{\varphi}} = \sin \theta.\label{eq:velocity}
}
Substituting these in the $r$ and $\eta$ formulae, we obtain
\dis{
&r=8\left(\frac{N_r}{D_r}\right)^2,
}
where
\dis{
&24N_r= \left[ 4 \lambda_{\phi} \phi (\phi^2 - f_{\rm DE}^2 ) - 4 \gamma \lambda_{X} \phi ( X^2 - \gamma (\phi^2 - m^2)) \right]
     \cos \theta  + \left[ 4 \lambda_X X (X^2 - \gamma (\phi^2 - m^2)) \right] \sin \theta,\\
     &24D_r=\lambda_{\phi} (\phi^2 - f_{\rm DE}^2)^2
     + \lambda_{X} (X^2 -  \gamma (\phi^2 - m^2 ))^2, \nonumber
}
and
\dis{
&\eta= \frac{{N}_\eta}{{D}_\eta} ,
}
where
\dis{
&{N}_\eta=  [ 4 \lambda_{\phi} (\phi^2 - f_{\rm DE}^2 ) + 8 \lambda_{\phi} \phi^2
     - 4 \lambda_X  \gamma (X^2 -  \gamma (\phi^2 - m^2 ) )   + 8 \lambda_{X}  \gamma ^2 \phi^2 ] \cos^2 \theta \\
     &\quad\quad+ 2 [
     - 8 \lambda_X  \gamma X \phi ( X^2   - \gamma (\phi^2 -m^2 ))  ] \cos \theta \sin \theta + [ 4 \lambda_X (X^2 -
      \gamma (\phi^2 - m^2))  + 8 \lambda_X X^2  ] \sin^2 \theta ,\\
     &{D}_\eta= \lambda_{\phi} (\phi^2 - f_{\rm DE}^2)^2
     + \lambda_{X} (X^2 - \gamma (\phi^2 - m^2 ))^2.\nonumber
}
Now let us suppose $\lambda_{X} = x \lambda_{\phi}$.  $X$ field rapidly converges to the ground state, in other words, the value of $X$ field is set to $X = \pm \sqrt{ \gamma (f_{\rm DE}^2 - m^2 ) }$. Then we can rewrite $r$ as
\dis{
r = \frac{128 \left( (1 +  \gamma ^2 x ) \phi \cos \theta - \gamma x \sin \theta \sqrt{ \gamma (f_{\rm DE} - m^2 ) }  \right)^2}{(1+ \gamma ^2 x)^2 (\phi^2 - f_{\rm DE}^2 )^2}.
}
Note that $\eta$ in this case is $\tilde{N}_\eta/\tilde{D}_\eta$ with
\dis{
\tilde{N}_\eta=&  4 ( 3 (1 +  \gamma ^2 x ) \phi^2 \cos^2 \theta - \gamma x (2 m^2 + \phi^2 ) \sin^2 \theta  + 4  \gamma ^2 x \phi^3 \cos \theta \sin \theta \sqrt{ \gamma (f_{\rm DE}^2 - m^2)} \\
& - f_{\rm DE}^2  (  (1 +  \gamma ^2 x ) \cos^2 \theta- 3  \gamma x \sin^2 \theta  + 4  \gamma ^2 x \phi \cos \theta \sin \theta \sqrt{ \gamma (f_{\rm DE}^2 - m^2 ) },\\
\tilde{D}_\eta=&(1+ \gamma ^2x) (\phi^2 - f_{\rm DE}^2 )^2.\nonumber
}

\begin{figure}[!t]
\begin{center}
\includegraphics[width=0.95\linewidth]{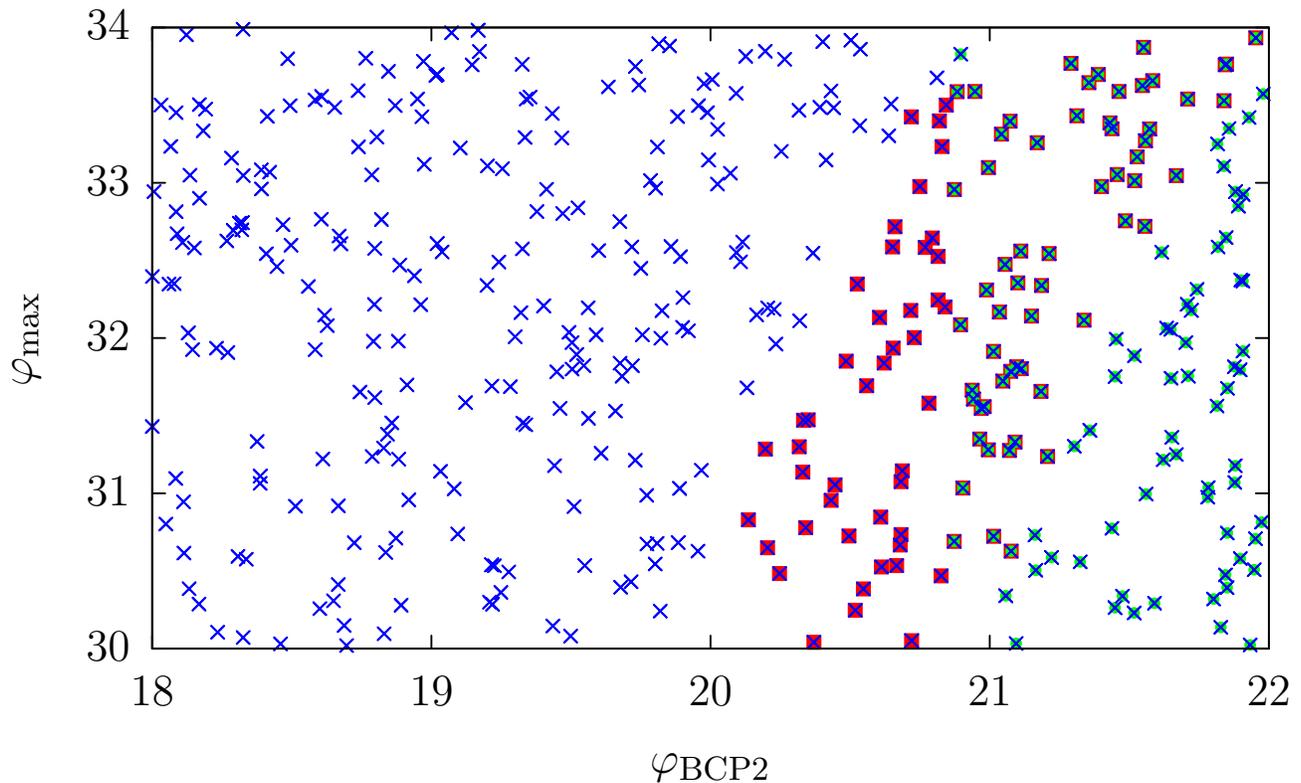}
\end{center}
\caption{The scatter plots for the $e$-fold number (red squares), $\eta$ (green bullets) and $n_s$ (blue crosses) as functions of $\varphi_{\rm init}$ and $\varphi_{\rm max}$. 
The points satisfying all these are shown as green bullets on top of the blue-cross and the red-square boundary.} \label{fig:scatter}
\end{figure}

\begin{widetext}
\begin{table}[!t]
\begin{center}
\begin{tabular}{cccc|ccc||ccc|c}
\hline
   $\lambda_\phi$& $\lambda_X$  &$\gamma$ &  $m$ & $\phi_{\rm BCP2}$&  $\phi_{\rm max}$ &$f_{DE}$ & $\eta$ &$ r$  & $n_s$ & $e$ \\
 \hline
  $0.620\times e^{-12} $&  $5.14\times e^{-12} $& 0.128&  0.0171& 21.84& 33.97&34.25& $3.26\times 10^{-3}$&    0.122&    0.961 & 55.44   \\
  $2.15\times e^{-13} $&   $0.740\times e^{-14} $&0.028&  0.0123&  21.53& 31.61&34.92 &$6.30\times 10^{-3}$ & 0.104& 0.974&  50.15 \\
 $0.726\times e^{-12} $&  $2.27\times e^{-15} $&  0.012& 0.0164& 28.20&39.98&41.56 &$7.31\times 10^{-3}$& 0.117&0.971& 50.35 \\
 $3.40\times e^{-14} $&  $    4.64\times e^{-13} $&  0.048& 0.0157&  29.03& 39.27& 40.37& $6.11\times 10^{-3}$& 0.152& 0.955&54.25 \\
\hline
\end{tabular}
\caption{The $e$-fold numbers for several parameter sets.}\label{tab1}
\end{center}
\end{table}
\end{widetext}

Note that the $x=0$ case is the same as the single field hilltop potential, \ie the vanilla hilltop. Since $x$ appears as a product with $\gamma^2$, \ie as $\gamma^2 x$. Though the deviation from the vanilla hiltop potential is controlled by both $\gamma$ and $x$, in our numerical study  it is sufficient to  examine the large $\gamma$ region only for a fixed $x$. The formulae are so complicated that it is not easy to realize the usefulness of the $X$ field toward a large
 $\eta$, but it turns out to be true. The procedure is the following. We introduced the velocity vector of   $\varphi$ in Eq. (\ref{eq:velocity}). The velocity vectors are basically obtained by solving the differential equation of $\varphi$ with the  assigned initial conditions. So we take arbitrary velocity vectors (corresponding to certain initial conditions) with the constraint $- \pi \leq \theta \leq \pi$, because $\varphi$ in the $\phi$ direction is monotonically increasing, namely, $\dot{\varphi}_{\phi} > 0$.  
 Now Eq. (\ref{eq:velocity}), through Eqs. (\ref{eq:onederi}) and (\ref{eq:twoderi}), gives directly $r$ and $\eta$ at the point from which CMB is originating, from which we conclude that the existence of an additional chaoton field is helpful for a large $\eta$. 
  
From Eqs. (\ref{eq:evolution}) and (\ref{eq:BC}), we determine $\phi(t)$ for $\phi_i>0$ and $\dot X(t_m)>0$ so that  $\varphi_{\rm BCP2}$ points of $r=[0.1,0.3]$ and $n_s=[0.94, 0.98]$ are obtained for given parameter sets of $\{f_{\rm DE},  \gamma , m, \lambda_\phi, \lambda_X  \}.$ The $e$-fold number is calculated numerically by
 \dis{
    e=\int_{\varphi_{\rm end}}^{\varphi_{\rm BCP2}}\,\frac{V(\varphi)}{V_{,\varphi}} d\varphi \,.
    }
But, the crucial dependence is summarized by the information on $\varphi_{\rm init}$  near  $\varphi_{\rm BCP2}$. So, the BC dependence on Eq. (\ref{eq:BC}) is to give an acceptable
$\varphi_{\rm init}$. The needed very small parameters are the quartic coupling constants, $\lambda_\phi=O(10^{-12})$ and $\lambda_X=O(10^{-13})$. In Table \ref{tab1}, the $e$-fold numbers for several sets of model parameters are shown. The decay constant is trans-Planckian of order O(30) as shown in Table \ref{tab1}.  In Fig. \ref{fig:scatter}, we present a scatter plot for some tried model parameters,
satisfying $r=[0.1,0.3],\, n_s=[0.94,0.98],$ and $e=[50,60]$.

\begin{figure}[!t]
\begin{center}
\includegraphics[width=0.95\linewidth]{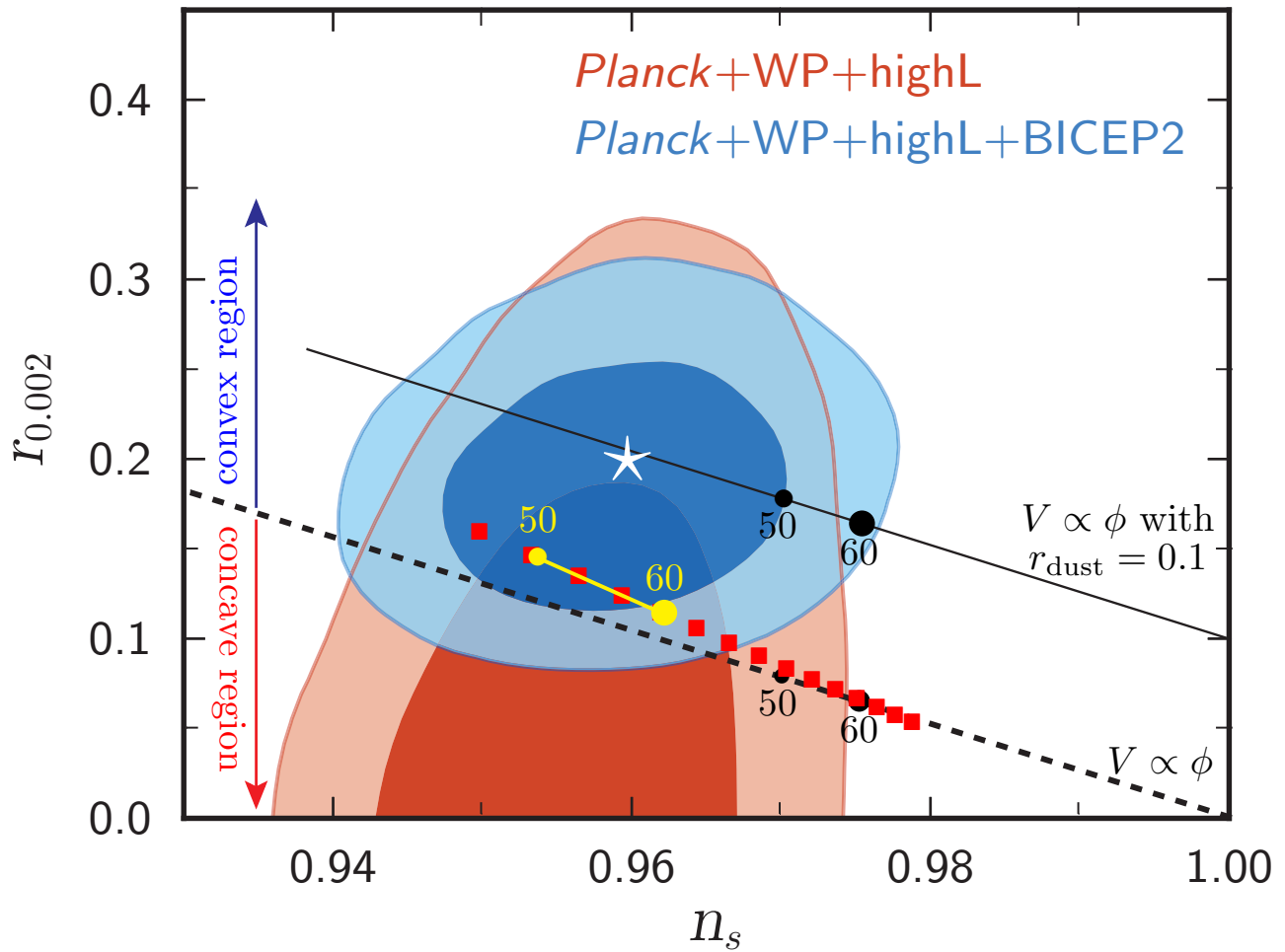}
\end{center}
\caption{The red boxes for the $e$-fold numbers around 55 with parameters of Eq. (\ref{eq:HilltopTwo}) from data points of Fig. \ref{fig:scatter} without the dust correction to $r$, \ie with $r_{\rm dust}=0$. The lines connecting $e=50$ and $e=60$ are sitting on top of these red squares, as shown with a yellow segment with two yellow bullets. The convex and concave potentials are separated by the thick black dash-line which is $V\propto \phi$, corresponding to  $r_{\rm dust}=0$. This linear potential including $r_{\rm dust}=0.1$ is shown as the black line. The star corresponds to  $n_s=0.96$ and $r=0.2$.} \label{fig:EFOLDING}
\end{figure}

In Fig. \ref{fig:EFOLDING}, we plot the $e$-fold numbers on the backgrounds of the Planck and BICEP2 data. The thick dash line separate the convex and concave potentials. The red squares are the points satisfying $e=[50,60]$. The segment satisfying $e=[50, 60]$ for a given set of model parameters is almost parallel to these red squares as shown with a yellow segment with two yellow bullets. 
  
\section{Conclusion}\label{sec:Conclusion}
We considered  a numerical calculation of the $e$-fold number, starting from a hilltop point with a hilltop inflationary potential of two inflatons, $\phi$ and $X$. The main inflaton is $\phi$, and $X$ is a sideway down-fall field in the region $\phi\ge m$ as schematically shown in Fig. \ref{fig:Path}.  By varying model parameters of Eq. (\ref{eq:HilltopTwo}) and initial conditions, the inflaton-path field $\varphi$ is moved (via the cosmological evolution equation of $\phi$ and $X$) toward the BICEP2 point,  $\varphi_{\rm BCP2}$ near $ n_s\simeq 0.96$ and $r\simeq 0.16$. Next, from the point  $\varphi_{\rm init}$, which is somewhat smaller than $\varphi_{\rm BCP2}$, to the end of inflation  $\varphi_{\rm max}$, we calculated numerically the $e$-fold number. In Fig. \ref{fig:scatter}, we presented the calculation as a scatter plot. In Fig. \ref{fig:EFOLDING}, we presented the $e$-fold number relation to $r$ and $n_s$ on the Planck and BICEP2 data backgrounds.  We find that a reasonable set of parameters are possible for the $e$-fold number in the range $50-60$. 
If the dust contribution $r_{\rm dust}$ is of order O(0.06), the red squares cross the star. 

\acknowledgments{J.E.K. is supported in part by the National Research
Foundation (NRF) grant funded by the Korean Government (MEST)
(No. 2005-0093841) and by the IBS (IBS-R017-D1-2014-a00), and  D.Y.M. is supported by BK21 Plus (No. 21A20131111123).}

\end{document}